\documentstyle[aps,prl,multicol,epsf,amssymb]{revtex}
\def\gs{\gtrsim}

\def\be{\begin{equation}}
\def\en{\end{equation}}                  
\newcommand{\bi}[1]{\mbox{\boldmath$#1$}}
\begin{document}
\draft
\bibliographystyle{prsty}
\title{Replica-exchange molecular dynamics simulation for supercooled liquids}
\author{Ryoichi Yamamoto$^{1}$ and Walter Kob$^{2}$}
\address{$^{1}$Department of Physics, Kyoto University, Kyoto 606-8502, 
Japan\\
$^{2}$Institute of Physics, Johannes-Gutenberg University, 
Staudinger Weg 7, D-55099, Germany}
\date{\today}
\maketitle

\begin{abstract}

We investigate to what extend the replica-exchange Monte Carlo method
is able to equilibrate a simple liquid in its supercooled state. We find
that this method does indeed allow to generate accurately the canonical
distribution function even at low temperatures and that its efficiency
is about 10-100 times higher than the usual canonical molecular dynamics
simulation.

\end{abstract}


\begin{multicols}{2}


If a liquid is cooled to a temperature close to its glass transition
temperature, its dynamical properties show a drastic slowing-down.
At the same time, a crossover from highly unharmonic liquid-like
behavior to harmonic solid-like behavior is expected in its static 
(thermodynamic) properties at a certain temperature $T_K$, 
the Kauzmann temperature \cite{Kauzmann}.
Very recently the value of $T_K$ of simple model
liquids have been determined analytically \cite{Mezard} and numerically
\cite{Sciortino,Coluzzi} and some possibilities of a thermodynamic
glass transition at $T_K$ have been discussed. Although the 
values of $T_K$ obtained with the different methods 
are consistent with each other, it was necessary for
the numerical calculations of $T_K$ to extrapolate high temperature
data ($T\gs0.45$) of the liquid and disordered solid branches of the
configurational entropy $S(T)$ down to significantly lower temperatures
($T_K\simeq0.3$). With a guide of an analytic prediction for liquids,
$S(T)\sim T^{-0.4}$ \cite{Rosenfeld}, and for harmonic solids, $S(T)\sim
\log T$, a crossing of the two branches has been found and used to
calculate $T_K$.  However, to make those observations more reliable,
very accurate calculations of thermodynamics properties are necessary
in the deeply supercooled regime, which is difficulat since the typical
relaxation times of the system are large.

In recent years, several efficient simulation algorithms have been
developed to generate canonical distributions also for complex systems.
Examples are the multi-canonical \cite{Berg,Lee},
the simulated tempering \cite{Lyubartsev,Marinari},
and the replica-exchange (RX) \cite{Hukushima,Swendsen} methods.
Although these methods were originally developed for Ising-type spin
systems, their applications to any off-lattice model by use of Monte
Carlo or molecular dynamics simulations are rather straightforward
\cite{Hansmann,Nakajima,Achenbach,Sugita}. However, it has been found
that the application of some of these algorithms to supercooled liquids
or structural glasses is of only limited use \cite{Bhattacharya}. The
main motivation of the present paper is to test the efficiency of the
RX method, which seems to be in many cases the most efficient algorithm,
to the case of highly supercooled liquids~\cite{Coluzzi2}.


The system we study is a two-component ($AB$) Lennard-Jones mixture,
which is a well characerized model system for supercooled simple
liquids.  The total number of particles is $N=1000$, and they interact
via the (truncated and shifted) 
potential $\phi_{\alpha\beta}(r_{ij})=4\epsilon_{\alpha\beta}
[(\sigma_{\alpha\beta}/r_{ij})^{12}-(\sigma_{\alpha\beta}/r_{ij})^6]$,
where $r_{ij}$ is the distance between particles $i$ and $j$, and the
interaction parameters are $\alpha,\beta\in A,B$, $\epsilon_{AA}=1$,
$\epsilon_{AB}=1.5$, $\epsilon_{BB}=0.5$, $\sigma_{AA}=1$,
$\sigma_{AB}=0.8$, and $\sigma_{BB}=0.88$. Other simulation parameters
and units are identical as in \cite{Kob}. The time step $\Delta t$
for numerical integration is $0.018$.

The algorithm of our replica-exchange molecular dynamics (RXMD) simulation
is essentially equivalent to that of Ref.~\cite{Sugita}, and therefore we
summarize our simulation procedure only briefly. (i) We construct a system
consisting of $M$ noninteracting subsystems (replicas), each composed of
$N$ particles, with a set of arbitrary particle configurations $\{{\bi
q}_1,\cdots,{\bi q}_M\}$ and momenta $\{{\bi p}_1,\cdots,{\bi p}_M\}$.
The Hamiltonian of the $m$-th subsystem is given by
\begin{equation}
H_m({\bi p}_m,{\bi q}_m)=K({\bi p}_m)+\Lambda_m E({\bi q_m}),
\end{equation}
where $K$ is the kinetic energy, $E$ is the potential energy, and
$\Lambda_m\in\{\lambda_1,\cdots,\lambda_M\}$ is a parameter to scale
the potential. (ii) A MD simulation is done for the total system, whose
Hamiltonian is given by ${\cal H}=\sum_{m=1}^M H_m$, at a constant
temperature $T=\beta_0^{-1}$ using the constraint method \cite{Allen}.
Step (ii) generates a canonical distribution $P({\bi q}_1,\cdots,{\bi
q}_M;\beta_0)= \Pi_{m=1}^M P({\bi q}_m;\Lambda_m\beta_0)\propto
\exp[-\beta_0\sum_{m=1}^M \Lambda_m E({\bi q_m})]$ in configuration space
\cite{Nose}.  (iii) At each time interval $\Delta t_{RX}$, the exchange
of the potential scaling parameter of the $m$-th and $n$-th subsystem
are considered, while $\{{\bi q}_1,\cdots,{\bi q}_M\}$ and $\{{\bi
p}_1,\cdots,{\bi p}_M\}$ are unchanged.  The acceptance of the exchange is
decided in such a way that it takes care of the condition of detailed balance.
Here we use the Metropolis scheme, and thus the acceptance ratio is given by

\begin{equation}
w_{m,n}= \left\{
\begin{array}{ll}
1,&\qquad \Delta_{m,n}\le 0\\
\exp(-\Delta_{m,n}),&\qquad\Delta_{m,n}> 0,
\end{array}
\right. 
\end{equation}
where $\Delta_{m,n}=\beta_0(\Lambda_n-\Lambda_m)(E({\bi q_m})-E({\bi
q_n}))$. (iv) Repeat steps (ii) and (iii) for a sufficient long
time. This scheme leads to canonical distribution functions $P(E;\beta_i)$
at a set of inverse temperatures $\beta_i=\lambda_i\beta_0$. To make a
measurement at an inverse temperature $\beta_l$ one has to average over
all those subsystems $(i\in 1,\cdots,M)$ for which we have (temporarily)
$\beta_l=\lambda_i\beta_0$. Usual canonical molecular dynamics (CMD)
simulations are realized if we skip step (iii).

In the present simulation, we take $M=16$, $\beta_0=0.45^{-1}$,
$\lambda_i=1-0.0367(i-1)$ and thus cover a temperature range $0.45\le T \le 1$.
Exchange events are examined only between subsystems that have scaling
parameters $\lambda_i$ and $\lambda_{i+1}$ that are {\it nearest}
neighbors; the events with $i=1,3,5,\cdots$ or $i=2,4,6,\cdots$ are
repeated alternatively every $\Delta t_{RX}$ intervals. We find that the
highest average acceptance ratio for this type of move is $0.186$ for
the exchange of $\lambda_1$ and $\lambda_2$, and the lowest is $0.027$
for $\lambda_{15}$ and $\lambda_{16}$. Although these values can be
made more similar by optimizing the different gaps between $\lambda_i$
and $\lambda_{i+1}$ for a fixed choice of $\lambda_1$ and $\lambda_M$,
only small improvements were obtained by such a simple optimization in our
case. We also note that the choice of $\Delta t_{RX}$ strongly affects
the efficiency of the RX method; $\Delta t_{RX}$ should be neither too
small or too large \cite{dtx}. We used $\Delta t_{RX}=10^3\Delta t$,
a time which is a bit larger than the one needed for a particle to do
one oscillation in its cage, and data are accumulated for $0\le t \le
5\times10^6\Delta t$ after having equilibrated the system for the same
amount of time. At the beginning of the production run, the subsystems
were renumbered so that at $t=0$ we had for all $m$ $\Lambda_m=\lambda_m$.


In Fig.~1(a), we show the time evolution of the subsystems in
temperature space. One can see that the subsystems starting from the lowest
($m=1$) and the highest ($m=16$) temperature explore both the whole
temperature space from $i=1$ to $16$. Fig.~1(b) presents the mean squared
displacements (MSD)
\begin{equation}
\Delta R^2(t)=|{\bi q}_m(t)-{\bi q}_m(0)|^2/N
\end{equation}
for the RXMD (with $m=1$) and for the CMD performed at
$T=(\lambda_1\beta_0)^{-1}=0.45$.  From this figure we recognize that,
due to the temperature variation in the RXMD method, the system moves
very efficiently in configuration space, while in the CMD the system is
trapped in a single metastable configuration for a very long time. If one
uses the MSD to calculate an effective diffusion constant, one finds that
this quantity is around 100 times larger in the case of the RXMD than
in the CMD case, thus demonstrating the efficiency of the former method.

Fig.~2 shows the canonical distribution function of the total 
potential energy at the different temperatures,
\begin{equation}
P_i(E)\equiv P(E;\lambda_i\beta_0),
\end{equation}
obtained by a single RXMD simulation. For adjacent temperatures
the corresponding distribution functions should
have enough overlap to obtain a reasonable
exchange probabilities and hence can be used to optimize the efficiency
of the algorithm. Further use of these distribution functions can be
made by using them to check whether or not one has indeed equilibrated the
system. 
Using the reweighting procedure \cite{Ferrenberg1},
it is in principle possible to calculate the canonical distribution functions 
\begin{equation}
P_i(E;\lambda_j\beta_0)=\frac{P_i(E)\exp[(\lambda_i-\lambda_j)\beta_0E]}
{\int dE' ~P_i(E')\exp[(\lambda_i-\lambda_j)\beta_0E']}
\end{equation}
at a new temperature $T_j=(\lambda_j\beta_0)^{-1}$ from any $P_i(E)$.
Note that in equilibrium the left hand side should be {\it independent}
of $i$ to within the accuracy of the data.

In Fig.~3 we plot different $P_i(E;\lambda_4\beta_0)$, using as input the
distributions $P(E;\lambda_i\beta_0)$ for $1\le i \le 8$, obtained from
RXMD (a) and CMD (b) simulations. (Both simulations extended over $8.7
\times 10^4$ time units.) We see that in the case of the RXMD
the different distributions $P_i$ fall nicely on top of each other in
the whole energy range, thus giving evidence that the system is indeed
in equilibrium. In contrast to this, the different distributions of
the CMD, Fig~3b, do not superimpose at low energies (=low temperatures), thus
demonstrating the lack of equilibration.
This can be seen more clealy by comparing Fig.~3(c) and (d),
where $P_i(E;\lambda_1\beta_0)$ is plotted.

Fig.~4 shows the temperature dependence of the potential energy $E(T)$
obtained from RXMD simulations via
\begin{equation}
E(T_j)=\int dE' ~P(E';\lambda_j\beta_0)E' \quad.
\end{equation}
For the sake of comparison we have also included in this plot data
from CMD with the same length of the production run as well as data
from CMD simulations which were significantly longer (about one order of
magnitude)~\cite{Sciortino}. The solid line is a fit to the RXMD results
with the function $E(T)=E_0+AT^{0.6}$, a functional form suggested by
analytical calculations~\cite{Rosenfeld}.  One can see that RXMD and CMD
results coincide at higher temperatures, but deviations become significant
at low temperatures (see Inset). Furthermore, we see that the present
RXMD results agree well with CMD data of the longer simulations.

As a final check to see whether the RXMD is indeed able to equilibrate
the system also at low temperatures, we have calculated the temperature
dependence of the (constant volume) heat capacity $C_v(T)$ via the
two routes
\begin{eqnarray}
C_v(T)&=&\partial E(T)/\partial T \\
&=&(\langle E^2\rangle-\langle E^2\rangle)/T^2,
\end{eqnarray}
and plot the results in Fig.~5. Again we see that within the accuracy
of our data the two expressions give the same answer, thus giving
evidence that the system is indeed in equilibrium.


Summary: We have done replica-exchange molecular dynamics and canonical
molecular dynamics simulations for a binary Lennard-Jones mixture
in order to check the efficiency of the replica-exchange method for
a structural glass former in the strongly supercooled regime. We find
that at low temperatures the RXMD is indeed significantly more efficient
than the CMD, in that the effective diffusion constant of the particles
is around 100 times larger in the RXMD. However, accurate simulations
are still difficult for $T<0.45$ even with RXMD. Finding an optimal
choice of $M$, $\{\lambda_1,\cdots,\lambda_M\}$, and $\Delta t_{RX}$
may be important in order to allow simulations also for $T<0.45$
within reasonable computation times. Furthermore it might be that the
efficiency of RXMD improves even more if one uses it below the critical
temperature of mode-coupling theory~\cite{mct}, since there is evidence
that below this temperature the nature of the energy landscape is not
changing anymore~\cite{horbach99}.


The authors acknowledge the financial support from the DFG through
SFB~262. RY acknowledges the Grants in Aid for Scientific Research from
the Ministry of Education, Science, Sports and Culture of Japan and
thanks Prof. B. Kim for valuable discussions. Calculations have been
performed at the Human Genome Center, Institute of Medical Science,
University of Tokyo.


\end{multicols}
\newpage

 \begin{figure}[t]
 \epsfxsize=2.5in
 \centerline{\epsfbox{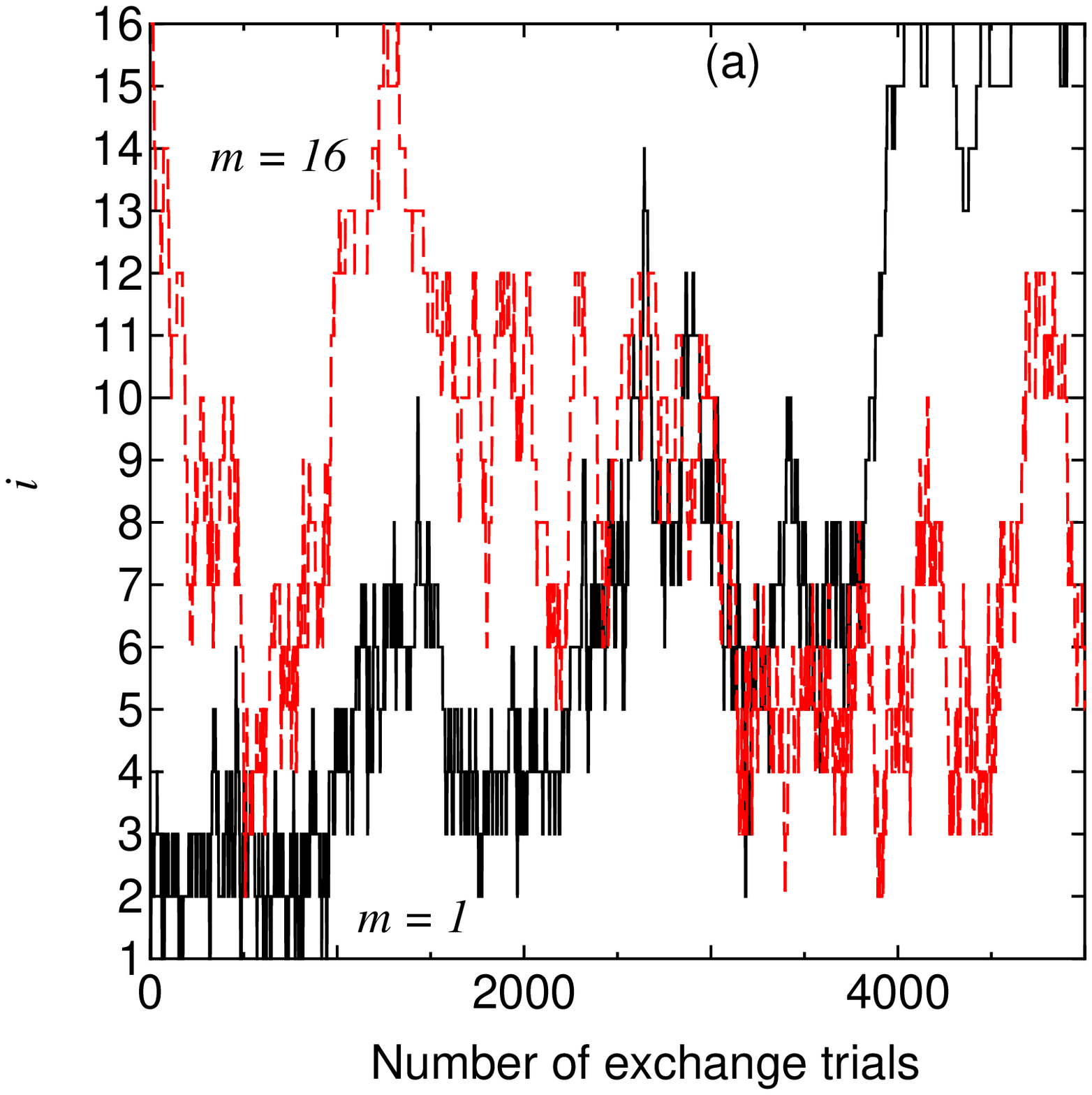}}
 \epsfxsize=2.6in
 \centerline{\epsfbox{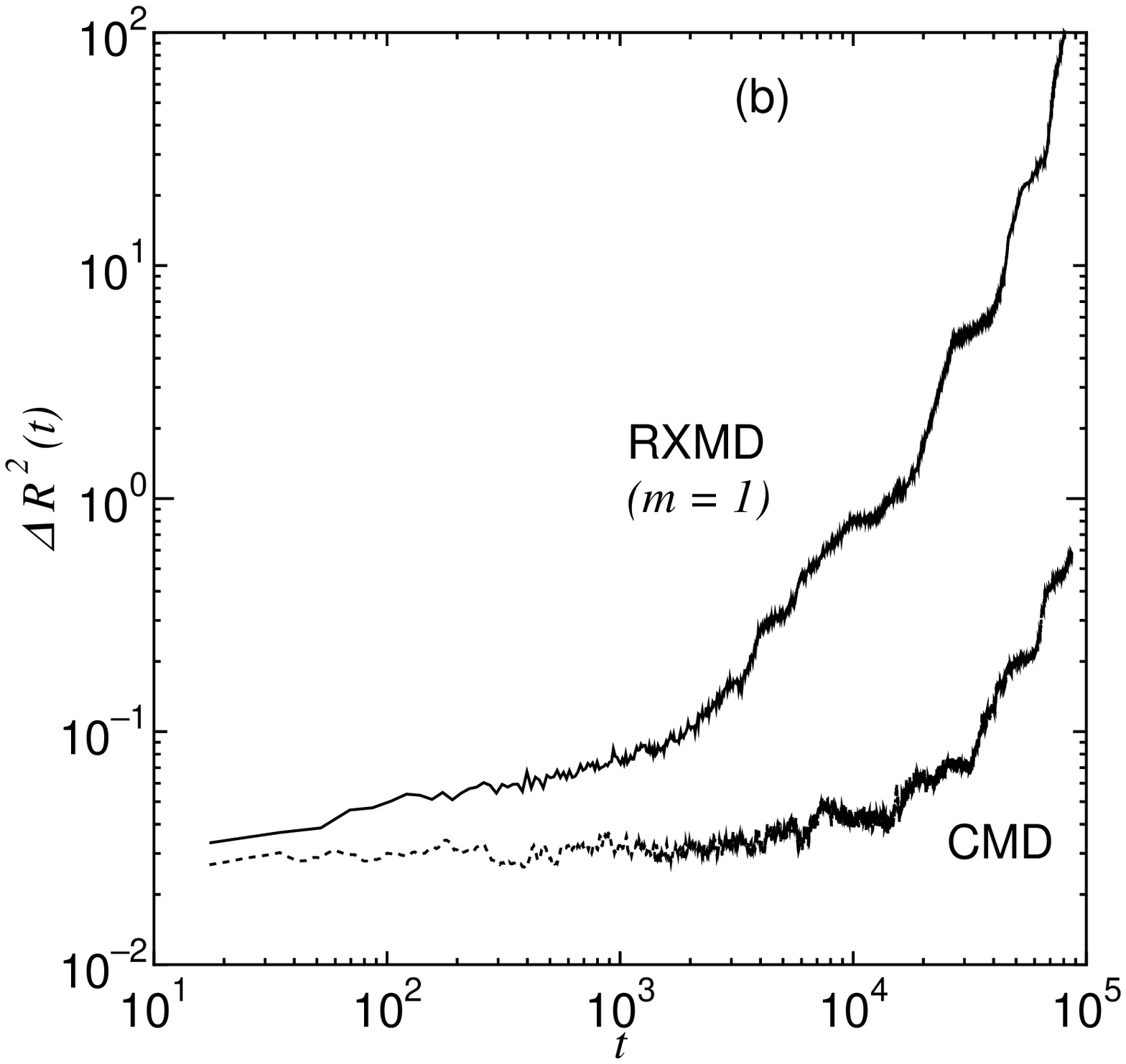}}
 \caption{\protect
(a) Typical walks of the subsystems in temperature space.
(b) Time dependence of the mean squared displacement. The solid line
shows $\Delta R^2(t)$ from RXMD for a subsystem which at $t=0$ was
at $T=0.45$ ($i=1$), and the dashed line is $\Delta R^2(t)$ from CMD
at $T=0.45$.  The two curves have been calculated by starting
from the same initial configuration.
}
 \label{fig1}
 \end{figure}

 \begin{figure}[t]
 \epsfxsize=2.6in
 \centerline{\epsfbox{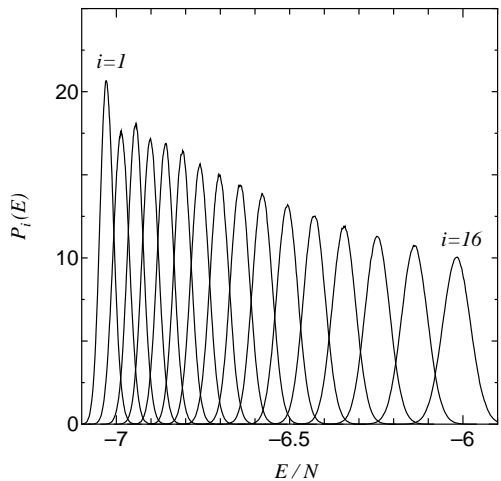}}
 \caption{\protect
The canonical distribution function $P_i(E)$ at various temperatures $T_i$ 
($1\le i\le16$) obtained by a single RXMD simulation. 
Here, $T_1=0.45$ and $T_{16}=1.0$.
 }
 \label{fig2}
 \end{figure}

\newpage
 \begin{figure}[b]
 \centerline{\epsfxsize=2.6in\epsfbox{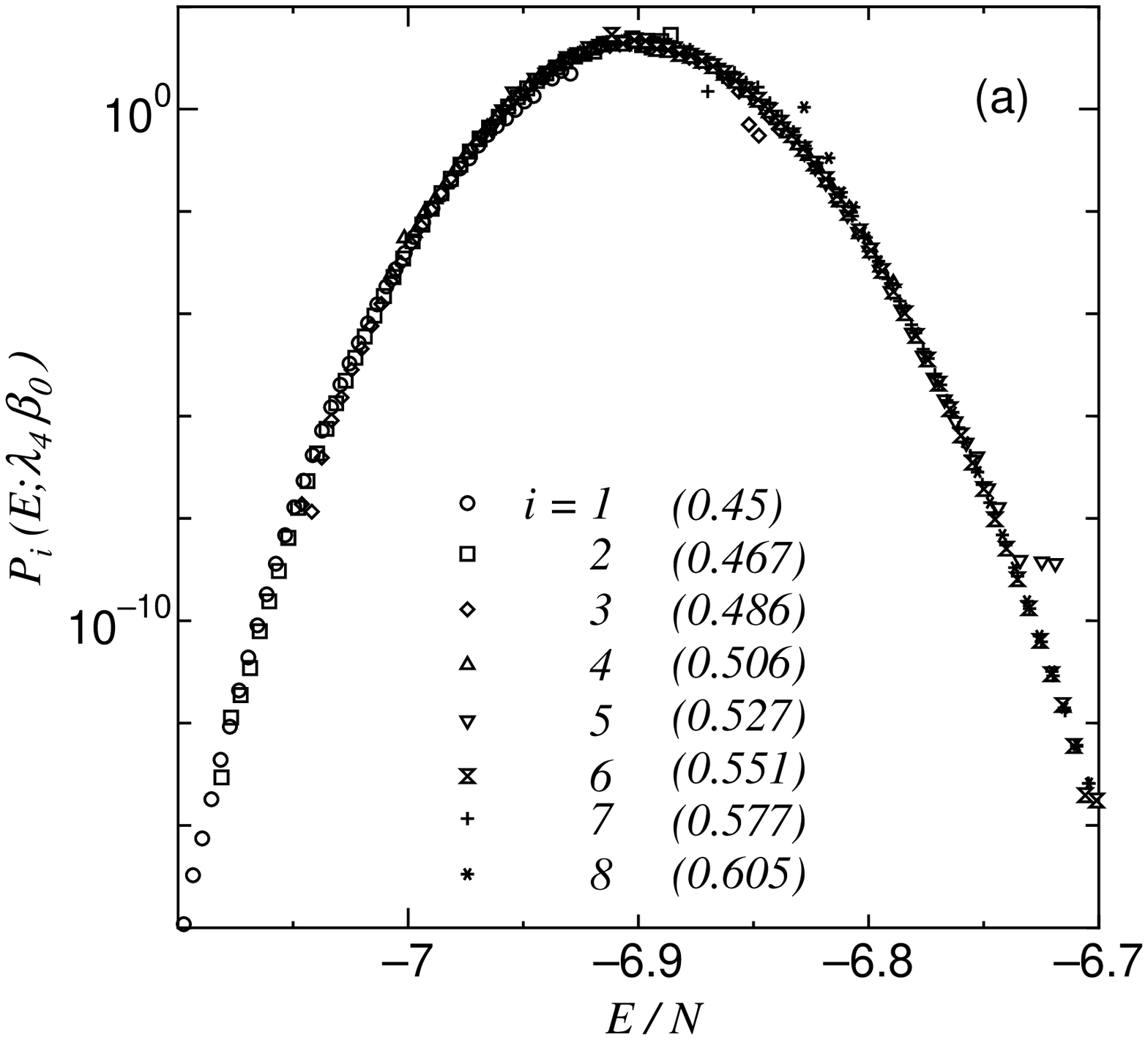}
             \epsfxsize=2.6in\epsfbox{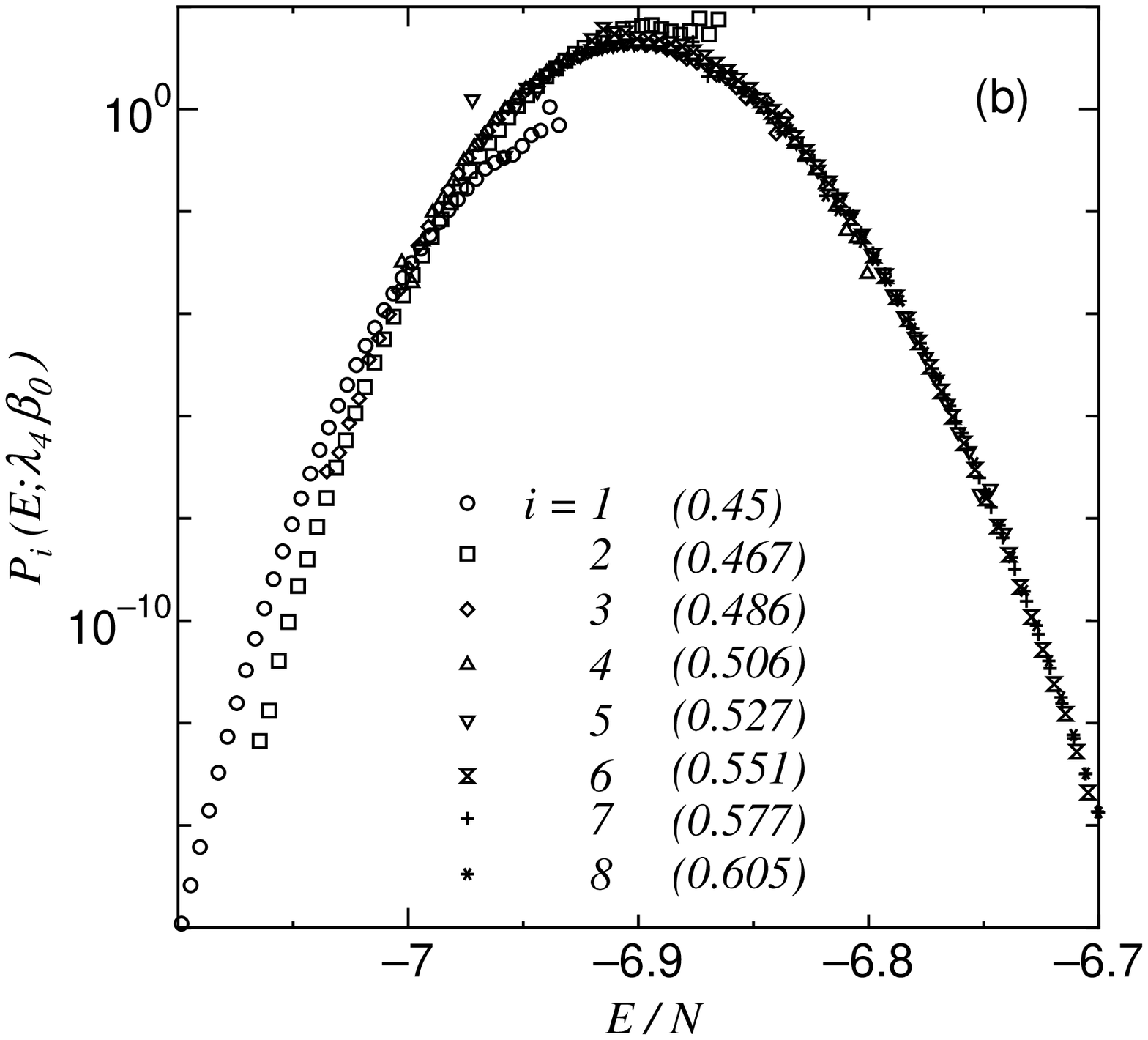}}
 \centerline{\epsfxsize=2.6in\epsfbox{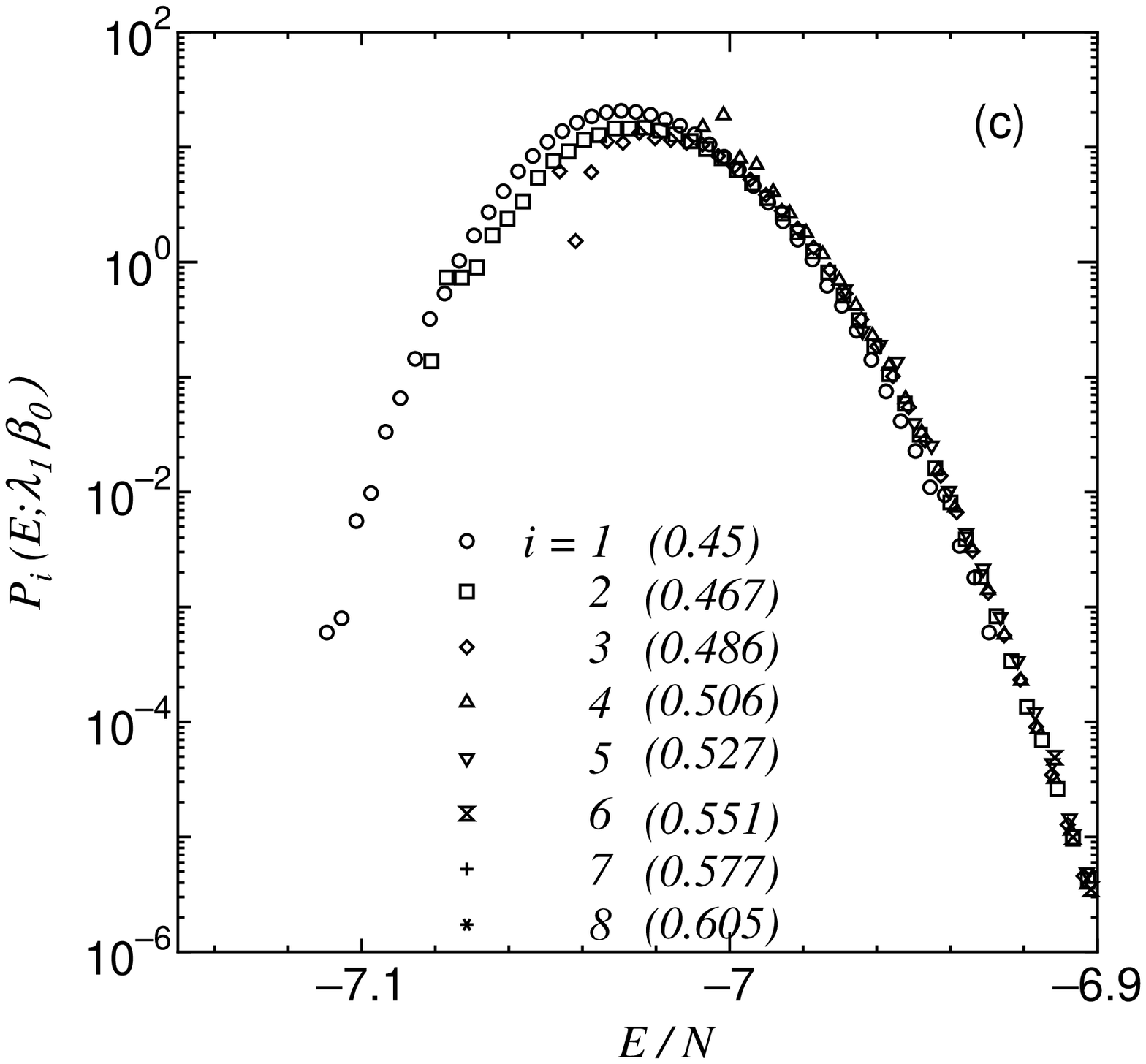}
             \epsfxsize=2.6in\epsfbox{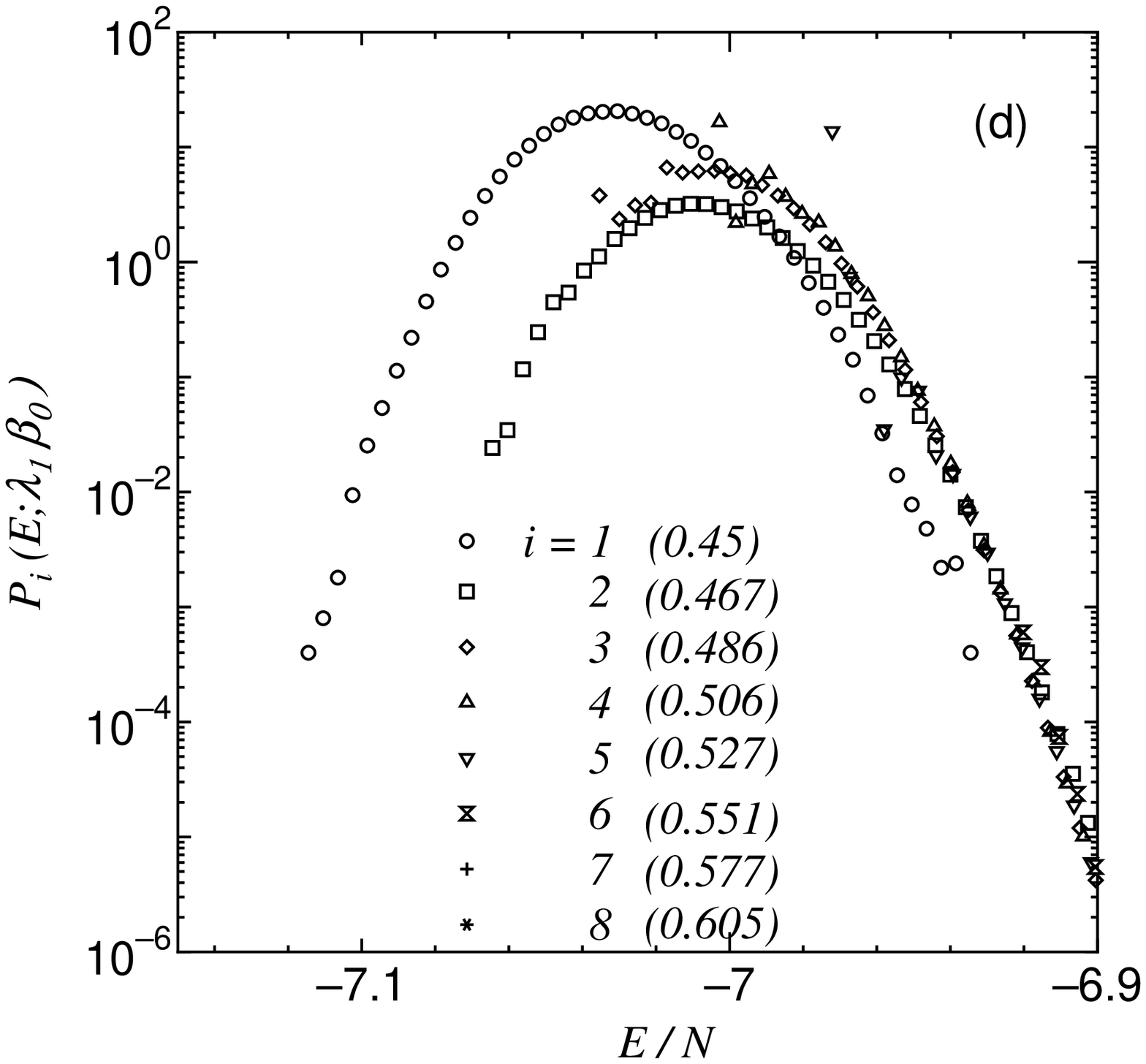}}
 \caption{\protect
The canonical distribution function at 
$T=(\lambda_4\beta_0)^{-1}=0.506$ by reweighting
$P_i(E)$ for $1\le i \le 8$ obtained by RXMD (a) and
standard CMD (b) simulations.
Numbers in the parentheses present temperatures at which simulations 
were done.
The same function at $T=(\lambda_1\beta_0)^{-1}=0.45$ 
obtained by RXMD (c) and CMD (d).
Note that in both simulations the length of
the runs is the same ($8.7\times10^4$ time units).
}
\label{fig3}
\end{figure}

 \begin{figure}[t]
 \epsfxsize=2.6in
 \centerline{\epsfbox{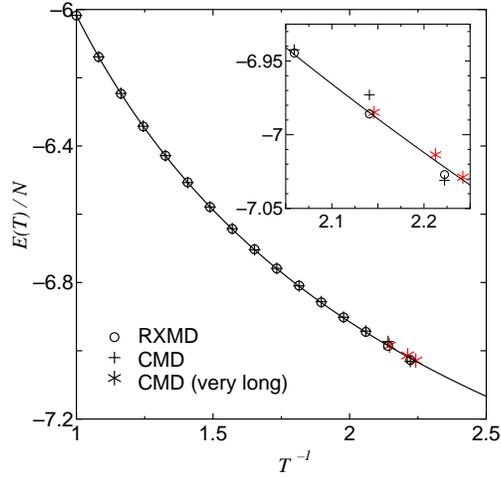}}
 \caption{\protect
Temperature dependence of the potential energy $E(T)$ obtained via
RXMD($\circ$) and CMD($+$) of runs with the same length. $\ast$ presents
values from much longer CMD runs. The solid line is the best fit to the
RXMD data with a fit function $E=E_0+AT^{0.6}$, where $E_0=-8.656$ 
and $A=2.639$ are fit parameters.
}
 \label{fig4}
 \end{figure}

 \begin{figure}[t]
 \epsfxsize=2.8in
 \centerline{\epsfbox{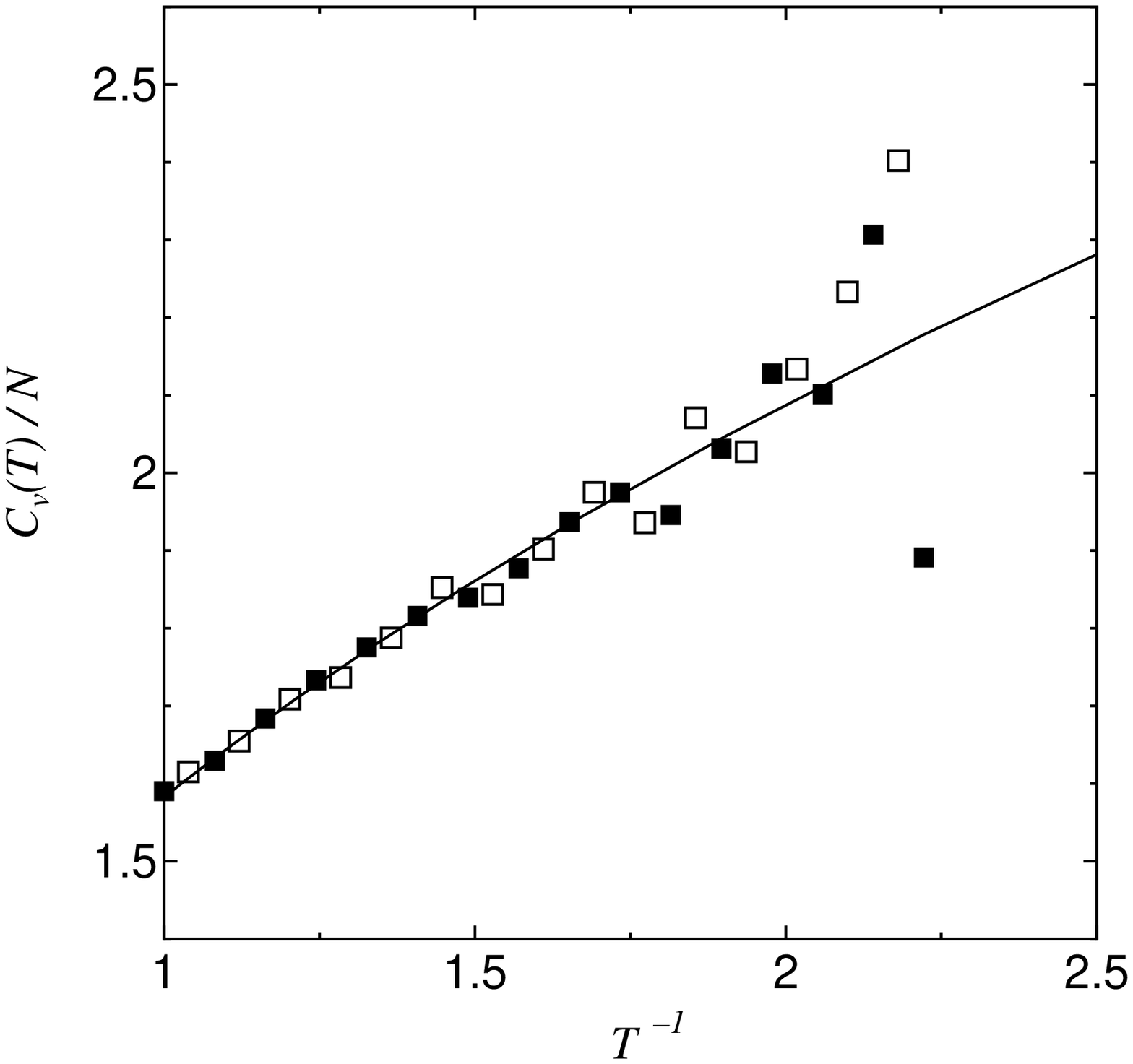}}
 \caption{\protect
Temperature dependence of the heat capacity $C_v(T)$ obtained via RXMD.
$\square$ presents data from $C_v=\partial E(T)/\partial T$, 
and $\blacksquare$ presents data from 
$C_v=(\langle E^2\rangle-\langle E\rangle^2)/T^2$.
The solid line is the result of a fit $C_v=0.6AT^{-0.4}$, with the
same value of $A$ as in Fig.~4.
 }
 \label{fig5}
 \end{figure}


\begin{references}
\bibitem{dtx}
If $\Delta t_{RX}$ is large, it obviously slows down the algorithm
since the number of exchange trials in a single simulation is inversely
proportional to $\Delta t_{RX}$. On the other hand, if an exchange
event between $\lambda_i$ and $\lambda_{i+1}$ (or $\lambda_{i-1}$)
has occurred at $t=0$, the properties of the subsystem at temperature
$\lambda_{i}\beta_0$ for $t \ge 0$ will depend for a characteristic
(aging or equilibration) time $\tau_m$ on whether it previously was at
temperature $\lambda_{i+1}\beta_0$ or $\lambda_{i-1}\beta_0$. So $\Delta
t_{RX}$ should be larger than $\tau_m$ for efficient random walks in the
temperature space, otherwise strong memory effects will make the exchanges
inefficient. The value of $\tau_m$ depends strongly on the details of the
system considered and is possibly very large in our system, particularly
at low temperatures. Actually with a choice of $\Delta t_{RX}=200\Delta
t$, the subsystems initially at the lowest three temperatures do not reach
the highest temperature state within our simulation time.
\end{references}
\end{document}